\documentclass[twocolumn,showpacs,superscriptaddress,amsmath,amssymb]{revtex4}
\usepackage{graphicx}
\usepackage{dcolumn}
\usepackage{bm}

\makeatletter

\newcommand{\Rmnum}[1]{\expandafter\@slowromancap\romannumeral #1@}
\makeatother

\def\bd{
\begin{document}} \def\ed{\end{document}}
\def\bmp{\begin{minipage}} \def\emp{\end{minipage}}
\def\bcc{\begin{center}} \def\ecc{\end{center}}     \def\npg{\newpage}
\def\beq{\begin{equation}} \def\eeq{\end{equation}} \def\hph{\hphantom}
\def\r#1{$^{[#1]}$}
\def\n{\noindent} \def\ni{\noindent} \def\pa{\parindent}
\def\hs{\hskip} \def\vs{\vskip} \def\hf{\hfill} \def\ej{\vfill\eject}
\def\cl{\centerline} \def\ob{\obeylines}  \def\ls{\leftskip}
\def\underbar#1{$\setbox0=\hbox{#1} \dp0=1.5pt \mathsurround=0pt
\underline{\box0}$}   \def\ub{\underbar}    \def\ul{\underline}
\def\f{\left} \def\g{\right} \def\e{{\rm e}} \def\o{\over} \def\d{{\rm d}}
\def\vf{\varphi} \def\pl{\partial} \def\cov{{\rm cov}} \def\ch{{\rm ch}}
\def\la{\langle} \def\ra{\rangle} \def\EE{e$^+$e$^-$} \def\pt{p_{\rm t}}
\def\pti{p_{\rm t,i}} \def\ptj{p_{\rm t,j}}
\def\bitz{\begin{itemize}} \def\eitz{\end{itemize}}
\def\btbl{\begin{tabular}} \def\etbl{\end{tabular}}
\def\btbb{\begin{tabbing}} \def\etbb{\end{tabbing}}
\def\beqar{\begin{eqnarray}} \def\eeqar{\end{eqnarray}}
\def\\{\hfill\break} \def\dit{\item{-}} \def\i{\item}
\def\bbb{\begin{thebibliography}{9} \itemsep=-1mm}
\def\ebb{\end{thebibliography}} \def\bb{\bibitem}
\def\bpic{\begin{picture}(260,240)} \def\epic{\end{picture}}
\def\akgt{\cl{\bf ACKNOWLEDGMENTS}}
\def\fgn{\noindent{\bf\large\bf figure captions}}
\def\lan{\langle}
\def\ran{\rangle}
\def\p{\pi}
\def\ifmath#1{\relax\ifmmode #1\else $#1$\fi}%
\def\rc{\ifmath{{\mathrm{c}}}}
\def\cut{\ifmath{{\mathrm{cut}}}}
\def\rF{\ifmath{{\mathrm{F}}}}
\def\rK{\ifmath{{\mathrm{K}}}}
\def\rp{\ifmath{{\mathrm{p}}}}
\def\rt{\ifmath{{\mathrm{t}}}}
\def\LAB{\ifmath{{\mathrm{LAB}}}}
\def\cut{\ifmath{{\mathrm{cut}}}}
\def\beq{\begin{equation}}
\def\eeq{\end{equation}}

\newcommand{\cinst}[2]{$^{\mathrm{#1}}$~#2\par}
\newcommand{\crefi}[1]{$^{\mathrm{#1}}$}
\newcommand{\crefii}[2]{$^{\mathrm{#1,#2}}$}
\newcommand{\crefiii}[3]{$^{\mathrm{#1,#2,#3}}$}
\newcommand{\HRule}{\rule{0.5\linewidth}{0.5mm}}

\bd
\title{Critical behavior of Binder ratios and ratios of higher order cumulants \\ of conserved charges in QCD deconfinement phase transition }

\author{Chen Lizhu}
\affiliation{Institute of Particle Physics, Hua-Zhong Normal
University, Wuhan 430079, China}\affiliation{Brookhaven National
Laboratory, Upton, NY 11973, USA}
\author{Pan Xue}
\affiliation{Institute of Particle Physics, Hua-Zhong Normal
University, Wuhan 430079, China}
\author{X. S. Chen} \affiliation{Institute of Theoretial
Physics, Chinese Academy of Sciences, Beijing 100190, China}
\author{Wu Yuanfang} \affiliation{Institute of Particle Physics,
Hua-Zhong Normal University, Wuhan 430079,
China}\affiliation{Brookhaven National Laboratory, Upton, NY 11973,
USA}\affiliation{Key Laboratory of Quark $\&$ Lepton Physics
(Huazhong Normal University), Ministry of Education, China }

\begin{abstract}
Binder liked ratios of baryon number are firstly suggested in
relativistic heavy ion collisions. Using 3D-Ising model, the
critical behavior of Binder ratios and ratios of higher order
cumulants of order parameter are fully presented. Binder ratio is
shown to be a step function of temperature. The critical point is
the intersection of the ratios of different system sizes between two
platforms. From low to high temperature through the critical point,
the ratios of third order cumulants change their values from
negative to positive in a valley shape, and ratios of fourth order
cumulants oscillate around zero. The normalized ratios, like the
Skewness and Kurtosis, do not diverge with correlation length, in
contrary with corresponding cumulants. Applications of these
characters in search critical point in relativistic heavy ion
collisions are discussed.
\end{abstract}


\maketitle

One of the main goals of current relativistic heavy ion experiments
is to locate the critical point of QCD deconfinement phase
transition. The critical character is that the correlation length
$\xi$ goes to infinite larger at infinite system. For finite system,
like the formed one in relativistic heavy ion collisions, the
correlation length should be a finite maximum. Therefore, the
various correlation length related observables are suggested in
relativistic heavy ion collisions~\cite{corr-fluc}.

It has been recently shown that near the critical point, the
density-density correlator of baryon-number follow the same power
law behavior as the correlator of the sigma field, which is
associated with the chiral order
parameter~\cite{stephanov,antoniou}. Therefore, the baryon number is
considered as an equivalent order parameter of formed system in
nuclear collisions~\cite{kapusta}.

From statistic physics, it also shows that the susceptibilities of
order parameter is directly related to the fluctuations of conserved
charges, i.e.,
\begin{equation}\label{susceptibility-1}
  \langle \delta N^i\rangle=VT\chi_i.
\end{equation}
\noindent $\chi_i$ is the $i$th order susceptibility. $\la\delta
N^i\ra=\la(N-\bar N)^i\rangle$ is the $i$th order cumulants of the
conserved charge number $N$. For three flavor QCD, the conserved
charges are baryon-number, strangeness, and electric
charge~\cite{koch}.

The third and forth order cumulants of conserved charges are defined
respectively as,
\begin{equation}\label{k34}
  K_3=\langle\delta N^3\rangle, \ \
  K_4=\langle\delta N^4\rangle-3\langle\delta N^2\rangle^2.
\end{equation}
\noindent In the vicinity of critical point, they are argued to be
proportional to the higher power of correlation length, i.e.,
$\xi^{4.5}$ and $\xi^7$~\cite{stephanov PRL,rajargopal},
respectively. So they are more sensitive to the correlation length,
and highly recommended.

In experiments~\cite{star-prl}, properly normalized cumulants, i.e.,
Skewness and Kurtosis,
\begin{equation}\label{kurtosis-ex}
  K_3/K_2^{3/2}=\frac{\langle\delta N^3\rangle}{\langle\delta
  N^2\rangle^{3/2}},\ \
  K_4/K_2^2=\frac{\langle\delta N^4\rangle}{\langle\delta
  N^2\rangle^2}-3,
\end{equation}
\noindent are actually presented. As the second cumulant is also
proportional to a certain power of correlation
length~\cite{non-monotonic}, if such normalized Skewness and
Kurtosis diverge with correlation length is not clear from the
theoretical point of view.

From theoretical side, the ratios of high order cumulants to the
second one, e.g.,
\begin{equation}\label{kurtosis-th}
  R_{3,2}=\frac{\la\delta N^3\ra}{\la\delta N^2\ra},
  \ R_{4,2}=\frac{\la\delta N^4\ra}{\la\delta N^2\ra}-3\la\delta N^2\ra.
\end{equation}
\noindent are estimated~\cite{karsch-prd,plb09,Liuyx,Skokov,Wuyl}. The Lattice
QCD with two light quark degrees of freedom shows that these ratios
of the baryon number, strangeness, and electric charge have
pronounced peaks from low to high temperature in the transition
region of chiral symmetry break~\cite{karsch-prd}. The effective
models in the mean-field approximation also shows that there are
peak, valley, and oscillating structures near the deconfinement and
chiral phase transitions~\cite{plb09,Liuyx,Wuyl}. However, all these
are obtained under some approximations due to the difficulties in
Lattice QCD calculations~\cite{Gupta} and model estimations~\cite{Skokov}.

Although the concrete form of interactions varies from one system to
another, according to the theory of university, the critical
exponent of equivalent measurement is identical in the same
university class. This allows us to study the critical behavior of
complex system by known simple one.

It is known that the QCD deconfinement phase transition corresponds
to the restoration of O(1) symmetry, which is the same university
class of 3D-Ising model~\cite{s-university}. Therefore, the critical
behavior of all above mentioned cumulants of baryon number can be
easily obtained from the corresponding cumulants of order parameter
in 3D-Ising model.

Moreover, it it known in statistical physics that the Binder ratio
of order parameter is a direct location of critical
point~\cite{binder}. Generally, the Binder liked ratios are
normalized raw moments of order parameter. The third and fourth
Binder liked cumulant ratios can be simply defined as,
\begin{equation}\label{binder-ratio}
  B_3=\frac{\langle M^{3}\rangle}{\langle M^{2}\rangle^{3/2}},
  \ \ B_4=\frac{\langle M^{4}\rangle}{\langle M^{2}\rangle^2}.
\end{equation}
\noindent Here we take 3D-Ising model as an example. The order
parameter in the model is the magnet $M=\sum_{i=1}^{N_L}
\vec{s_i}/N_L$ of spin $\vec{s}$ in all lattice sites $N_L$.

Equivalently, the order parameter in relativistic heavy ion
collisions is the baryon number. The temperature, or the controlling
parameter, is the incident energy. The size of the formed system is
mainly determined by the overlapped area, i.e., centralities. So if
we pass through the region of critical incident energies in
relativistic heavy ion collisions, the Binder ratios of baryon
number can be served as a good location of critical point of QCD
deconfinement phase transition.

In this paper, we firstly present the critical behavior of Binder
ratios in 3D-Ising model, and demonstrate why they are helpful, in
particular, in locating the critical point in relativistic heavy ion
collisions. Then, the critical behavior of Skewness, Kurtosis,
$R_{3,2}$, and $R_{4,2}$ are presented and discussed, respectively.
Meanwhile, from finite-size scaling of the susceptibilities, the
critical behavior of those ratios are estimated model independently.
Finally, the conclusions are drawn.

The critical behavior of Binder ratios, $B_3$ and $B_4$, in 3D-Ising
for 4 different lattice sizes are presented in Fig.~1(a) and (b),
respectively. Where the simulation of 3D-Ising model is based on the
wolff algorithm~\cite{MCbook}. We can see that both $B_3$ and $B_4$
show a step jump in the vicinity of critical temperature. The
physical meaning of this jump is clear.

When the temperature is much lower than the critical one, the system
is almost order and the fluctuation of order parameter is very
small, i.e.,
\begin{equation}\label{order-phase}
  \langle M^n\rangle\sim\langle M\rangle^n
  \;\;\;\;\;\;\;({\rm for}\  n=2,3,4\cdots).
\end{equation}
\noindent So it results the lower platform, which is 1 for all
orders of Binder ratios at all system sizes, as shown in Fig.~1.
When the temperature approaches to critical one, the correlation
length starts to increase with temperature and the fluctuations
become larger and larger. Their critical behavior is system size
dependent and described by finite-size scaling. Only at critical
temperature, all size curves intersect to the {\it fixed point},
where they are system size independent~\cite{fs1}, as shown in
Fig.~1. When the temperature is much higher than the critical one,
the system is totally disordered. It approaches again to a constant.
This forms the platform at high temperature. It is 1.6 and 3 times
larger than the lower platforms for the third and fourth order Binder
ratios, respectively. So the higher the order of Binder ratio, the
larger the gap of the step function.

\begin{figure}
\includegraphics[width=3.5in]{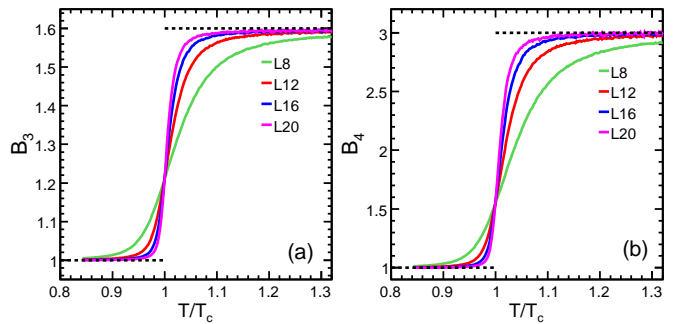}
\caption{\label{Fig. 1}(Color online) The temperature dependence of
Binder ratios in Eq.~(\ref{binder-ratio}) in the vicinity of
critical temperature in 3D-Ising model for 4 different lattice
sizes.}
\end{figure}

This step function liked behavior can be served as a very good probe
of critical point in relativistic heavy ion collisions, where
critical incident energy is difficult to assign precisely in priori.
So if we scan incident energies, and observe two platforms at low
and high energy regions, respectively, then the critical one is most
probably between them. We can finely tune the incident energy in the
region and precisely determine the critical energy and exponents.

The Skewness and Kurtosis of order parameter in 3D-Ising model for 4
different lattice sizes are presented in Fig.~2(a) and (b),
respectively. We can see from the figure that they change sharply in
the vicinity of the critical temperature. The Skewness first drops
down and then goes up, and Kurtosis oscillates with temperature.
Their values are system size dependent. Their signs change
respectively near the critical point. Former in Fig.~2(a) changes
from negative to positive when the temperature is increased through
the critical point, while the later in Fig.~2(b) becomes negative
only when the temperature is close to the critical point. The sign
change in Skewness, or third order cumulants, is expected in
effective models~\cite{Asakawa,Liuyx,Wuyl}.

\begin{figure}
\includegraphics[width=3.5in]{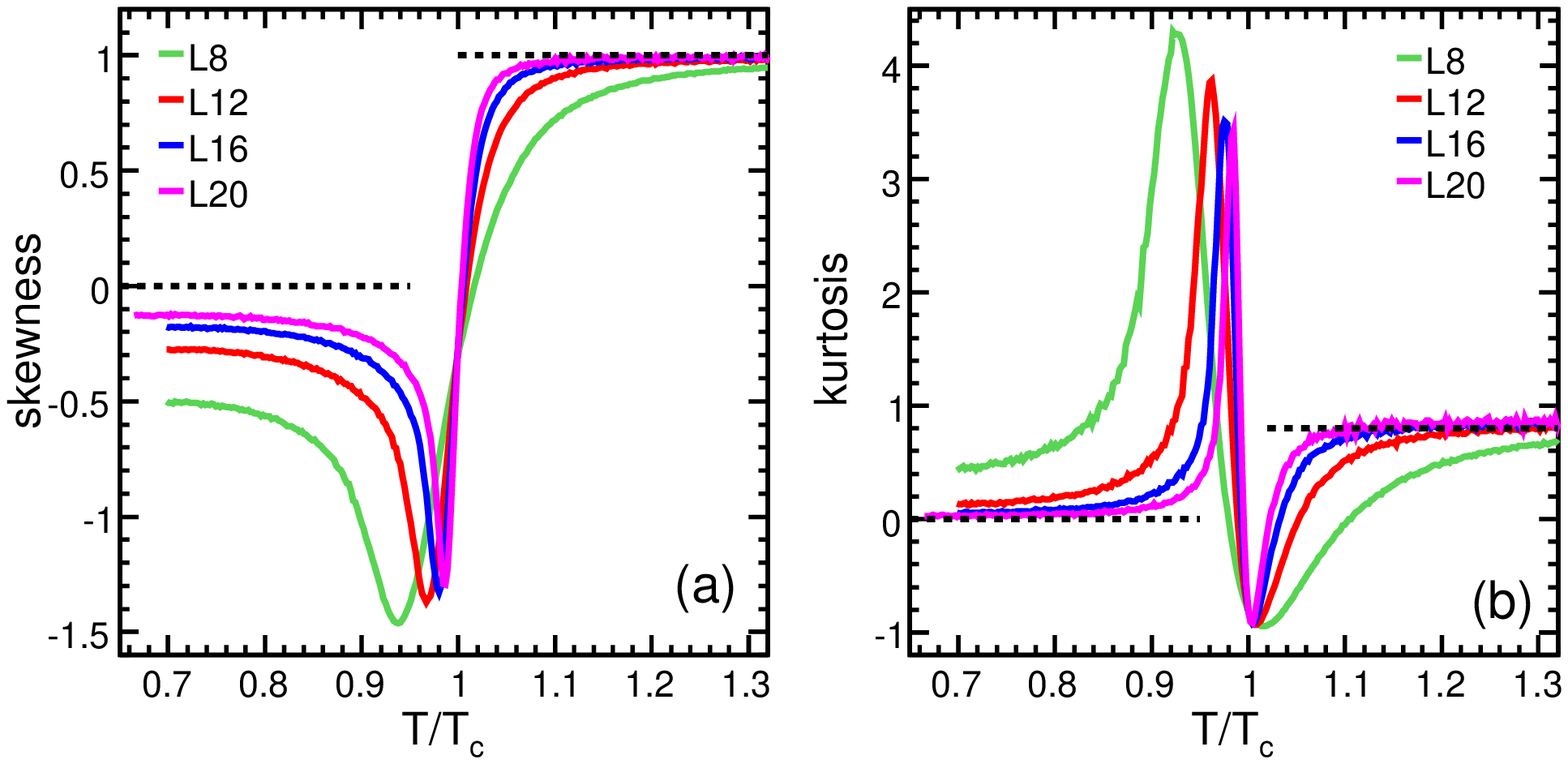}
\caption{\label{Fig. 2}(Color online) The temperature dependence of
Kurtosis (a) and Skewness (b) in Eq.~(\ref{kurtosis-ex}) in the
vicinity of critical temperature in 3D Ising model for 4 different
lattice sizes.}
\end{figure}

As we know that the Skewness and Kurtosis measure the symmetry and
sharpness of the distribution, respectively. The distributions of
order parameter $M$ near the critical point at system size $L=8$ are
shown in Fig.~3. Where we can clearly see that the long tail of the
distributions changes from the left to the right side when the
temperature is increased through the critical point, and the peak of
the distribution vary from sharp to flat when temperature is
approached the critical point. The same trend has been observed in
percolation model, in studying clusterization phenomena in nuclear
multi-fragmentation~\cite{percolation}.

\begin{figure}
\includegraphics[width=3.5in]{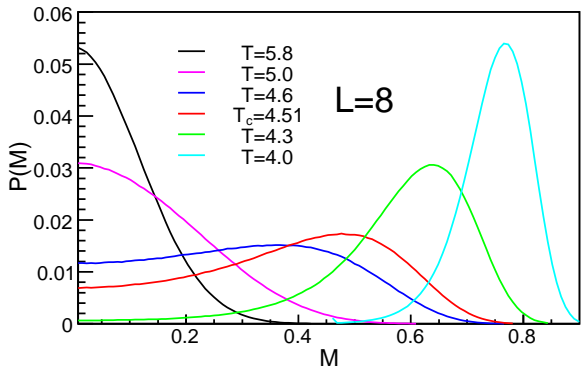}
\caption{\label{Fig. 3}(Color online) The distributions of order
parameter near critical temperatures in 3D-Ising model at system
size $L=8$.}
\end{figure}

This character can also be served as a signal associated with the
appearance of critical point in relativistic heavy ion collisions.
If we observe sign change of Skewness (Kurtosis) of baryon number at
a certain incident energy region, it most probably predicts the
appearance of critical point in the nearby incident energy region.

The Skewness and Kurtosis also converge to two constants when the
temperature is away from critical point, as shown in Fig.~2(a) and
(b). But the constants at low and high temperatures are close to
zero and 1, respectively. The gap between them are small and does
not change very much with the order of cumulants, unlike the Binder
ratio.

Moreover, all size curves of Skewness (Kurtosis) intersect at
critical point. This can be easily understood from finite-size
scaling of susceptibilities, i.e., \beqar
\chi_i(t,L)=L^{\gamma_i/\nu}P_{\chi_i}(t L^{1/\nu}).\eeqar\noindent
Where the $\gamma_i$ is the critical exponents of $i$th order
susceptibility, and $\nu$ is the critical exponent of correlation
length $\xi_{\infty} = t^{-\nu}$ at infinite system.
$t=\frac{T-T_{\rm c}}{T_{\rm c}}$ is reduced temperature, and
$T_{\rm c}$ is critical temperature. In the vicinity of critical
point, the correlation length at finite system is approximately the
same order of the system size, i.e., $\xi\sim L=V^{1/3}$.

For $\chi_3$ and $\chi_4$, $\gamma_{\rm 3}/\nu=4.5$, $\gamma_{\rm
4}/\nu=7$, respectively~\cite{stephanov PRL}. So the critical
behavior of the Skewness and Kurtosis in Eq.~(\ref{kurtosis-ex})
are, \beqar\label{fs-skewness}
 K_3/K_2^{3/2} &=&\frac{VT\chi_3}{(VT)^{3/2}\chi_2^{3/2}}\sim
\frac{L^3L^{4.5}P_{\chi_3}(t L^{1/\nu})}{L^{4.5}L^3T^{1/2}P_{\chi_2}^{3/2}(t L^{1/\nu})}\nonumber \\
 &=& T^{-1/2}F_S(t L^{1/\nu}),\nonumber\\
 K_4/K_2^2 &=& \frac{VT\chi_4}{(VT)^2\chi_2^2} \sim \frac{L^3L^7P_{\chi_4}(t L^{1/\nu})}{L^6L^4TP_{\chi_2}^2}\nonumber \\
 &=&T^{-1}F_K(t L^{1/\nu}).\eeqar
\noindent They no long diverge with correlation length, or system
size. At the critical temperature $t=0$, the scaling function, i.e.,
$F_S(0)$ or $F_K(0)$, is system size independent constant. All size
curves intersect to the constant, i.e., the {\it fixed
point}~\cite{fs1}.

From this simple estimation and Fig.~2, we can see that normalized
high order cumulants, i.e., Skewness and Kurtosis, do not directly
diverge with correlation length any more, different from
corresponding cumulants, $K_3$ and $K_4$, which are proportional to
$\xi^{4.5}$ and $\xi^7$, respectively~\cite{stephanov
PRL,rajargopal}.

The $R_{3,2}$, and $R_{4,2}$ of order parameter in 3D-Ising model
for 4 different lattice sizes are presented in Fig.~4(a) and (b),
respectively. We can see again from Fig.~4(a) that $R_{3,2}$ changes
its value sharply from negative to positive when temperature is
increased through the critical point. $R_{4,2}$ in Fig.~4(b)
oscillates greatly with temperature near the critical point. These
qualitative features, i.e., sign change in third moment, and
oscillating structure in forth cumulants, are consistent with
estimations of effective models~\cite{Asakawa,Liuyx,Wuyl}.

\begin{figure}
\includegraphics[width=3.5in]{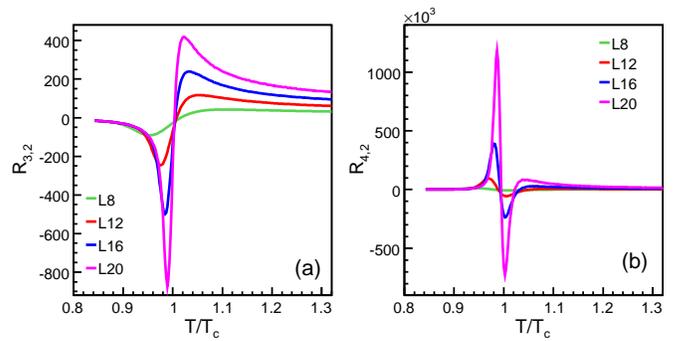}
\caption{\label{Fig. 4}(Color online) The temperature dependence of
$R_{3,2}$ (a), and $R_{4,2}$ (b) in the vicinity of critical
temperature in 3D-Ising model for 4 different lattice sizes.}
\end{figure}

$R_{3,2}$ and $R_{4,2}$ are very sensitive to the system size, or
correlation length. Their values become very large when system size
increases. The critical exponent of $R_{3,2}$, and $R_{4,2}$ can be
roughly estimated from finite-size scaling of susceptibilities,
i.e., \beqar\label{R42-fixpoint}
R_{3,2}&=&\frac{VT\chi_3}{VT\chi_2}=
  \frac{L^3\xi^{4.5}P_{\chi_3}(t L^{1/\nu})}{L^3\xi^2P_{\chi_2}(t
  L^{1/\nu})}\nonumber\\
  &=& \xi^{2.5}F_{R_{3,2}}(t L^{1/\nu})\nonumber\\
  R_{4,2} &=& \frac{VT\chi_4}{VT\chi_2} =
  \frac{L^3\xi^7P_{\chi_4}(t L^{1/\nu})}{L^3\xi^2P_{\chi_2}(t
  L^{1/\nu})}\nonumber\\
  &=& \xi^{5}F_{R_{4,2}}(t L^{1/\nu}).
\eeqar \noindent So $R_{3,2}$ and $R_{4,2}$ diverge with correlation
length as $\xi^{2.5}$ and $\xi^5$, respectively.

In this paper, the measurements of Binder liked ratios of conserved
charges are firstly suggested in relativistic heavy ion collisions.
Using 3D-Ising model, it is shown that near the critical
temperature, Binder ratios is a step function of temperature. The
gap of the step function is 1.6 and 3 times wider for the third and
forth order Binder ratios, respectively. This can be served
as a good identification of critical behavior in relativistic heavy
ion collisions, where the critical incident energy is unknown in
prior. The critical point is the intersection of Binder ratios of
different size systems between two platforms.

The critical behavior of Skewness, Kurtosis, $R_{3,2}$ and $R_{4,2}$
at various system sizes are also studied by 3D-Ising model, and
estimated by finite size scaling. When the temperature is increased
through the critical point, the ratios of the third order cumulants
change their values from negative to positive in a valley shape, and
ratios of fourth order cumulants oscillate around zero. All size
curves of Skewness (Kurtosis) intersect at the critical point. The
normalized ratios, like the Skewness and Kurtosis, do not diverge
with correlation length. While, un-normalized ratios, $R_{3,2}$ and
$R_{4,2}$, are divergent with correlation length. They are
proportional to $\xi^{2.5}$ and $\xi^5$, respectively, and very
sensitive to the system size near the critical temperature.

These critical characters may show up at the energy dependence of
corresponding ratios of conserved charges. Their behavior at coming
relativistic heavy ion experiments at RHIC, SPS, and FAIR are called
for.

\skip 0.5cm

We are grateful for stimulated discussions with Dr. Nu Xu. The first
and last authors are grateful for the hospitality of BNL STAR group.
This work is supported in part by the NSFC of China with project No.
10835005 and MOE of China with project No. IRT0624 and No. B08033.

\ed